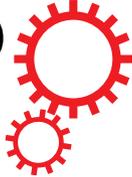

# OPEN  How costly punishment, diversity, and density of connectivity influence cooperation in a biological network

Ivan C. Ezeigbo[1,2]

It has been an old unsolved puzzle to evolutionary theorists on which mechanisms would increase large-scale cooperation in human societies. Thus, how such mechanisms operate in a biological network is still not well understood. This study addresses these questions with empirical evidence from agent-based models designed to understand these network interactions. Repeated Prisoner's Dilemma games were designed to study how costly punishment, diversity, and density of connectivity interact to influence cooperation in a biological network. There were 1000 rounds in each game made up of 18 players engaged in pairwise relationship with their neighbors. This study shows three important interactions. (1) Introducing diversity to costly punishment favors both cooperation and defection, but not vice versa. Introducing costly punishment to diversity disfavors defection but favors cooperation. (2) Costly Punishment, alone, disfavors defection but decreases average payoff. Decreasing the density of connectivity, $D_c$, when there is no costly punishment applied, increases average payoff. (3) A synergy of diversity and decreasing density of connectivity favors cooperation in a biological network. Furthermore, this study also suggests a likelihood from empirical findings that spatial structures may not be favoring cooperation, as is the widely-accepted notion, but rather disfavoring defection in the global scale.

Evolutionary theorists have long pondered the mechanisms that favor emergence of cooperation in a biological network[1–4]. Some studies have shown that costly punishment could favor the emergence of cooperation[5–8], whereas other studies have rather shown that cooperation is not always favored[9]. Cooperation is important because it sustains everyone in the network, as opposed to defection, which sustains the defector alone. In fact, it is an established fact that the evolutionary success of our species is largely because of our evolved capacity to cooperate[10–12]. Hence, defectors certainly obtain higher net payoffs than cooperators but it is obvious that a network having solely defectors is not sustainable. Punishment has long been suspected to be a potential 'curative' to defection; however sometimes, this punishment could be costly[13,14]. Costly punishment is simply paying a cost, $\alpha$, to detract a payoff of $\beta$ from another player. Designing human experiments to understand these interactions has advantages, but also constraints. Some of those constraints are (1) social and cultural bias could extensively impact experimental results. There have been studies that show that the emergence of cooperation or adoption of an evolutionary mechanism is partly affected by the societal or cultural characteristics[9,15–17]. Also, people who hold strong ties to each other are more likely to cooperate than defect[18]. Thus, there is more room for bias with human experiments. (2) The researcher has little control over variables and may not be able to freely isolate the effect of certain parameters. (3) There is a limitation on data size, a very common constraint with human experiments. Large enough data size is needed to counteract misleading effects of randomness. These constraints have led to different conclusions in the literature on the influence of punishment on cooperation[7,19] and on the influence other network features, like diversity, have on cooperation for different games and populations[20,21]. Biological networks are complex systems, and counteracting or reducing these three constraints would give us a clear image as to how these mechanisms play in the system and recurring patterns that govern these interactions.

[1]College of Computational Science, Minerva Schools at KGI, California, USA. [2]College of Natural Science, Minerva Schools at KGI, California, USA. Correspondence and requests for materials should be addressed to I.C.E. (email: ivyezeigbo@yahoo.com)

| Part A | | | |
|---|---|---|---|
| | **Opponent gets** | | |
| **Player gets** | **C** | −c | ±b |
| | **D** | ±d | −d |
| | **P** | −α | −β |
| **Part B** | | | |
| | **C** | **D** | **P** |
| **C** | b−c | −d−c | −β−c |
| **D** | b+d | 0 | −β+d |
| **P** | b−α | −d−α | −β−α |
| **Part C** | | | |
| | **C** | **D** | **P** |
| **C** | 2 | −2 | −5 |
| **D** | 4 | 0 | −3 |
| **P** | 2 | −2 | −5 |

**Table 1.** Payoff Matrix for Prisoner's Dilemma Game. The underlined table inputs give payoff of networks where costly punishment is not implemented. C are cooperators, D are defectors and P are costly punishers.

To circumvent these constraints, an agent-based model of repeated Prisoner's Dilemma games was designed in this study to understand the interactions in biological networks. I designed 100 games and each game had 1,000 rounds. This appeared to be the optimal number of rounds in a game since increasing the number of rounds in a game from 1000 does not affect the result. 18 players were used for the results presented in this paper, and the number of neighbors, k, varied randomly per player. Players are represented as nodes in a graph, and their connections with other players are represented as edges. In every round, each player interacts in a pairwise Prisoner's Dilemma game with his neighbors. In other words, if Player A is connected to $k$ other players in a particular round of a game, then Player A would perform $k$ different pairwise Prisoner's Dilemma games in that round. Just like in Gomez-Gardenes et al[22], before each round, each player considers the payoff of its neighbors (other players connected to that player) and copies the strategy of the neighbor with the highest payoff, given that it is higher than his. A game would end under two conditions (i) if all players assume the same strategy, or (ii) after 1000 rounds, there still exist more than one surviving strategy in the network, in which case, the game is said to be in an equilibrium/ 'No Preference' state. Since 1000 is the optimal number of rounds after which almost no change is observed, a network existing in an equilibrium/ 'No Preference' state remains in this state even if the number of rounds is increased from 1000 – almost no change would be observed. The bar chart produced after the simulation is run gives the number of times (the probability) a strategy emerges as the sole surviving strategy of the system in all 100 games, or how many times a round is completed without all players assuming the same strategy - the equilibrium/ 'No Preference' state. The payoffs were calculated following Dreber et al[23]. Cooperation meant paying 1 unit for the other player to gain 3 units. Defection meant gaining 1 unit at the cost of 1 unit for the other person and costly punishment meant paying 1 unit for the other person to lose 4 units. Table 1 shows the payoff matrix.

The density of connectivity, $D_c$, is simply the total number of edges in the connected graph. This is a very important feature that is, unfortunately, not well explored in the literature. The density of connectivity was varied to understand its impact in a biological network. In varying the density of connectivity, the number of nodes/players, N, in the network was multiplied by some variable, f. That is,

$$D_c = N * f \ (f \text{ is a multiplying factor}) \quad (1)$$

Given that we must always have a connected graph, $D_c$ is varied within the range of $N-1$ (which is the least density of connectivity) to $\binom{N}{2}$, which is the total number of edges for a fully connected graph (the highest density of connectivity). Since N is a constant, the multiplying factor, f is the only parameter that needs to be varied for any given N. We can use equation (1) to determine the range of $f$:

For the least density of connectivity, we have:

$$\begin{aligned} D_c &= N * f = N - 1 \\ f &= \frac{N-1}{N} \\ f &= \frac{N-1}{N} \text{ (value of f for the least } D_c \text{ possible)} \end{aligned} \quad (2)$$

For the highest density of connectivity, we have:

$$D_c = \binom{N}{2} = \frac{N(N-1)}{2} = N\left(\frac{N-1}{2}\right)$$

$$\text{If } N\left(\frac{N-1}{2}\right) = N*f \Rightarrow f = \frac{N-1}{2}$$

$$f = \left(\frac{N-1}{2}\right) \text{ (value of f for the highest } D_c \text{ possible)} \quad (3)$$

Hence, from equation (2) and equation (3), we have demonstrated that $f$ falls within the range, $\frac{N-1}{N} \leq f \leq \frac{N-1}{2}$.

The unique relevance of this study, as it concerns addressing the question of evolutionary theorists on cooperation, is that with the power of larger data or sample size, we can observe how evolution occurs not just at the local scale, but also at the global scale. Therefore, we can draw probabilistic predictions on the evolution of cooperation or defection in the network when certain parameters are changed. In this study, the parameters manipulated to understand interactions in a complex biological network were diversity, density of connectivity, and costly punishment. This study does two important things. Firstly, it suggests three interactions that occur at the global scale of a biological network, and secondly, it suggests that spatial structures have a similar network interaction as costly punishment. This study suggests that the following three interesting interactions occur at the global scale in a biological network: (1) Introducing diversity to costly punishment favors both cooperation and defection, but not vice versa. However, introducing costly punishment to diversity disfavors defection but favors cooperation. (2) Costly Punishment, alone, disfavors defection (but may not necessarily favor cooperation, as discussed below); nonetheless, it decreases average payoff. There are empirical and theoretical studies that show evidence that punishment is effective against selfishness in groups where most members are defectors[12,24]. Decreasing the density of connectivity, $D_c$, when there is no costly punishment applied, increases average payoff. Decreasing $D_c$ also reduces the probability of the dominant strategy in a network, as is demonstrated in this study. In a fully connected graph, defection is the dominant strategy, hence decreasing $D_c$ would disfavor defection and allow emergence of some cooperators in the network. Studies by other researchers that is consistent with this result show that limiting the amount of interaction among agents would allow the survival of cooperators, as these cooperators tend to form clusters that protects them from exploitation or invasion by defectors[25]. (3) A synergy of diversity and decreasing $D_c$ favors cooperation. Many studies give support to the influence of diversity on the emergence of cooperation[25–29]. In this study, we show that cooperation is even further amplified as $D_c$ decreases. Unlike at the local scale, which is the level of analysis of many previous experimental work on human cooperation in biological networks, the evolutionary strategy is not binary at the global scale. That is, at the local scale, a player may either cooperate or defect; however, at the global scale, we could observe cooperation, defection, or the equilibrium/ 'No Preference' state (where at least two strategies coexist). Hence, disfavoring one strategy, such as defection, would mean favoring the other strategy, cooperation, at the local scale since it is binary. However, the same logic can be deceptive at the global scale, since disfavoring one strategy, such as defection, would not necessarily mean that cooperation is "solely" favored. Disfavoring defection at the global scale, whether resulting in an equilibrium state or a solely cooperative strategy, invariably means that more cooperators are in the network, but not necessarily that cooperation is favored. In other words, the fact that cooperation is favored at the local scale does not mean that it is also favored at the global scale. Therefore, introducing costly punishment alone in a network would disfavor defection but may not necessarily solely favor cooperation. For this reason, it can be misleading to draw conclusions from associations in complex systems observed only at the local scale, as opposed to both the global and local scale. Gracia-Lazaro et al[20] carried out a human experimental study showing that players defected more than they cooperated when playing a Prisoner's Dilemma game in a heterogeneous and diverse network. On replicating their experiment on square lattices, my study suggests that this is true at the local scale. However, at the global scale, this study shows an interesting scenario of nonlinearity where cooperation emerges as the more likely sole strategy in a square lattice spatial structure with diversity. Nonlinearity is a typical feature of complex systems where the whole is more than the sum of its parts. In the replication of this experiment, we observe a very similar pattern to the interaction we see when diversity is introduced to costly punishment. With spatial structures, introducing diversity also favors both cooperation and defection at the global scale. However, since we have seen that costly punishment introduced alone interacts in a network by disfavoring defection but not necessarily solely favoring cooperation in the global scale, this study proposes the possibility that spatial structures at the global scale also disfavor defection. This is consistent with the widely-accepted notion that spatial structures favor cooperation[30,31]; although at the global scale, as proposed in this study, we say it rather disfavors defection.

### Results

My results suggest the following important relationships in a repeated Prisoner's Dilemma game:

### Introducing diversity to costly punishment favors both cooperation and defection, but not vice versa. Introducing costly punishment to diversity disfavors defection but favors cooperation.

Diversity is implemented by randomizing the connection of players in the network and how many people a player is connected to in each round of a game. Figure 1 shows the structure of a random network created by the agent-based model, with an f of 1.5. Figure 2, however, shows the relationship of costly punishment, diversity, and both costly punishment and diversity in a network with an f of 1.5. This demonstrates the interaction we observe on introducing diversity to costly punishment as opposed to introducing costly punishment to diversity.

For f = 1.5

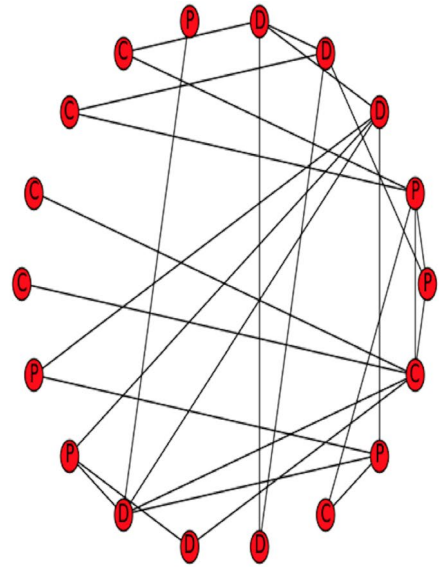
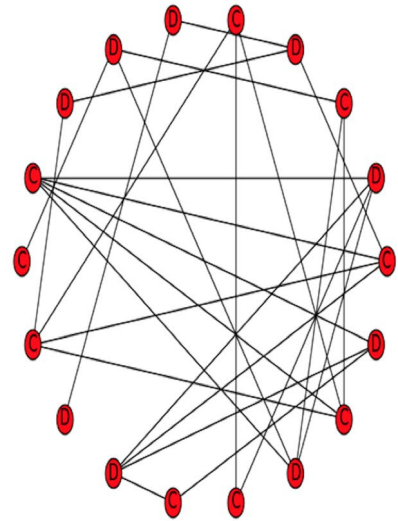

Network with punishers (with costly punishment)

Network with no punisher (no costly punishment)

**Figure 1.** Structure of the Biological Network For N = 18, f must be in the range $\frac{N-1}{N} \leq f \leq \frac{N-1}{2}$.

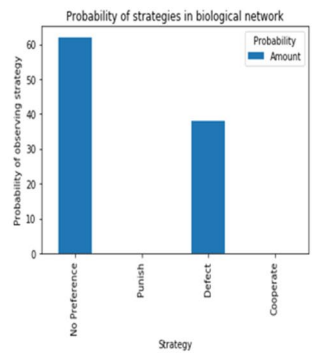

Counter({'No Preference': 62, 'Defect': 38, 'Punish': 0, 'Cooperate': 0})

Costly Punishment without Diversity in Network

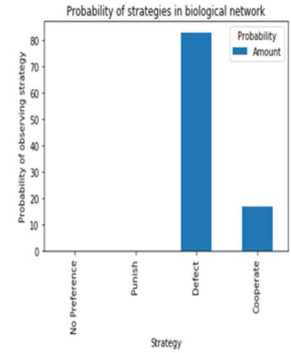

Counter({'Defect': 83, 'Cooperate': 17, 'No Preference': 0, 'Punish': 0})

Costly Punishment + Diversity in Network

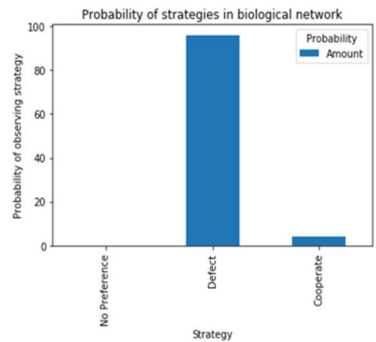

Counter({'Defect': 96, 'Cooperate': 4, 'No Preference': 0})

Diversity without Costly Punishment in Network

**Figure 2.** Diversity and Costly Punishment in Biological Network, when f = 1.5.

D are defectors, P are punishers, and C are cooperators. Graph plotted with NetworkX Python Package. Defector, cooperator or punisher assignments to nodes were done randomly. The degree and connections of nodes were also randomized, using only the density of connectivity. Equal number of cooperators, punishers, and defectors (that is, six of each) were assigned to the left panel while equal number of cooperators and defectors (nine of each) were assigned to the right panel.

**Costly Punishment, alone, disfavors defection but decreases average payoff.** Results from the model show that at any given value of f, introducing costly punishment alone (without diversity) decreases defection and favors an equilibrium/ 'No Preference' state. Thus, this goes further to demonstrate why 'disfavoring defection' does not necessarily mean 'favoring cooperation' at the global scale. However, since defection gives the highest payoff, introducing costly punishment alone comes at a price - average payoff decreases significantly. Dreber et al[23] also reported a similar consequence of costly punishment on average payoff. With lower values of f, we would also expect lower average payoffs; however, this is not the case without costly punishment. Decreasing the density of connectivity, $D_c$, when there is no costly punishment applied, increases the average payoff. Superficially, this is a surprising result because decreasing $D_c$ when there is no costly punishment causes a decrease in defection, the dominant strategy, and favors the equilibrium/ 'No Preference' state just in the same way as decreasing $D_c$ when costly punishment is applied. On closer inspection however, we find that this is not surprising at all. The results from the agent-based model show that the equilibrium state for both are different. The equilibrium state for the network without costly punishment consist of only defectors and thriving cooperators with high payoffs that compete just as well as these defectors; whereas the equilibrium state with costly punishment consist of defectors and punishers, with defectors having low payoffs and punishers having even much lower payoffs from the costly punishment. Figure 3 shows us how defection and average payoff changes when costly punishment is introduced for two networks – one with an f of 2.5 and the other with an f of 1.1. Both networks in Fig. 3 also show how decreasing $D_c$ decreases defection, the dominant strategy, in favor of the equilibrium state.

**A synergy of diversity and decreasing density of connectivity favors cooperation in a biological network.** From results from the computational model, we see that decreasing f, gradually, from $\frac{N-1}{2}$ towards $\frac{N-1}{N}$ decreases the dominant strategy and converts this to an equilibrium/ 'No Preference' state. The dominant strategy is usually defection, since in a fully connected network with no punishment, defection would be the sole surviving strategy as it gives the highest payoff and all other nodes would assume the strategy of the neighbor with the highest payoff[32]. In fact, we would only need one mutant defector in the network to make defection emerge as the sole surviving evolutionary strategy for any number of N. This makes sense if we consider the expected value (E.V.) of the payoffs for cooperators and defectors.

Let's take $N_c$ to to be the total number of cooperators, and $N_d$ to be the total number of defectors. In a fully connected network, every cooperator, $C_i$, is connected to $N_c - 1$ other cooperators and $N_d$ defectors, and every defector, $D_j$, is connected to $N_d - 1$ other defectors and $N_c$ cooperators. Then, expected value of payoffs using the payoff matrix in Fig. 1 is given below.

$$\begin{aligned} E.V._{C_i} &= N_c - 1 \ (cooperator - cooperator \ payoff) + N_d(cooperator - defector \ payoff) \\ &= N_c - 1 \ (2) + N_d(-2) = 2N_c - 2 - 2N_d \end{aligned} \quad (4)$$

$$\begin{aligned} E.V._{D_j} &= (N_d - 1)(defector - defector \ payoff) + N_c \ (defector - cooperator \ payoff) \\ &= (N_d - 1)(0) + N_c(4) = 4N_c \end{aligned} \quad (5)$$

Equation (4) and equation (5) show us that the expected value for every single cooperator $E.V._{C_i}$ is less than half the expected value of every defector, $E.V._{D_j}$ even if we take $N_d$ to be equal to 1 (that is, one mutant defector). Diversity plays a role by converting this equilibrium/ 'No Preference' state to cooperation. Diversity alone cannot favor cooperation since at a fairly high value of f, inducing diversity does not really help or change anything since the network is almost fully connected or homogeneous. In a similar vein, decreasing density of connectivity cannot on its own favor cooperation as well, but will only favor cooperation in synergy with diversity which would convert the 'No Preference' state to cooperation. Figure 4 shows this relationship when there are no punishers for $N = 18$.

Like the result shows, since an *f* of 2.4 still gives a network that is almost fully connected (homogenous), diversity does not influence cooperation as there is no equilibrium state to transform to the 'Cooperate' strategy so do not observe cooperation when we introduce diversity. Hence, diversity cannot influence cooperation on its own. If we increase the number of games to 10,000 games though, we would be able to observe some cooperation with a very probability (less than 1%). With much higher values of *f*, this becomes less likely, and defection would remain the sole strategy. However, with much lower values of *f*, defection is disfavored for an increase in the equilibrium/ 'No Preference' state. When diversity is introduced, this equilibrium/ 'No Preference' state is transformed to cooperation. If costly punishment is introduced, we observe something different since we know that costly punishment when introduced alone favors the equilibrium state over defection. This would mean that we would have an enhanced equilibrium state with decreasing $D_c$. But this equilibrium state is not one which can be directly transformed to cooperation when diversity is introduced because, as discussed in (II) above, this equilibrium state consists of 'punishers and defectors' rather than 'cooperators and defectors'. However, we expect to see an enhancement of cooperation due to the synergy of costly punishment and diversity; we should also expect

**a) When f = 2.5**

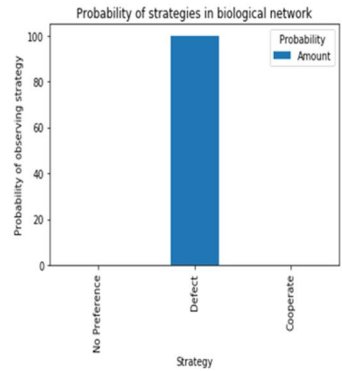 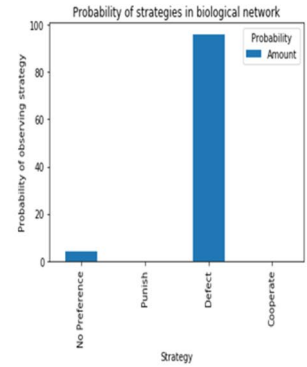

**Without** costly punishment          **With** costly punishment

**b) When f = 1.1**

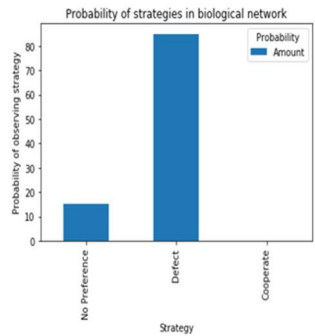 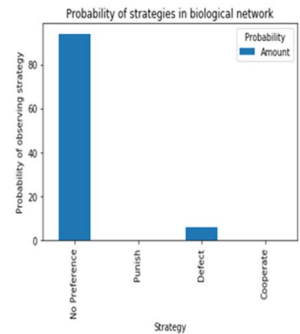

**Without** costly punishment          **With** costly punishment defection to be favored with introduction of diversity to costly punishment as discussed in (I). Figure 5 shows the interactions we observe when we introduce costly punishers into the network for the same value of *N*.

**Figure 3.** Costly Punishment and Average Payoff For N = 18, f must be in the range $\frac{17}{18} \leq f \leq \frac{17}{2}$ (**a**) When f = 2.5. (**b**) When f = 1.1.

### Discussion

Gracia-Lazaro *et al*[20] conducted experiments with square lattices and heterogeneous networks to conclude that heterogeneous networks do not favor cooperation when humans play Prisoner's Dilemma games. This study however shows that diversity or heterogeneity in a network favors cooperation but only in synergy with decreasing density of connectivity. There are studies in other areas of research where some diversity in a network have contributed to the promotion of cooperation in Prisoner's Dilemma games[33,34]. I created an agent-based model of a 25 × 25 square lattice, with 625 players, to investigate these rather conflicting results just like they did. I kept the number of neighbors constant at k = 4, but varied the players at every round of the game. In this case, this was the experimental treatment group; the control treatment group did not have variations with the players in each round. The only major differences were (1) the payoff matrix was calculated following Dreber *et al*[23] rather than Grujic *et al*[35] (2) they conducted human experiments; whereas I used computational agent-based models (3) They looked at rounds ranging from 50 to 70 of 1 game for both networks they analyzed while I looked at 100 games, each having 1000 rounds. Players copy the strategy of connected players in their network, and a game would stop if all players have the same strategy or 1000 rounds were completed. Updates are synchronous and the game begins initially with 313 defectors and 312 cooperators. Since this is an agent-based model, rather than a human experiment, the problem of bias with having finite number of rounds does not apply. My results from replicating their first experiment with the square lattice rather support my findings that diversity in synergy with decreasing density of connectivity promotes cooperation in a network. Figure 6 shows this result.

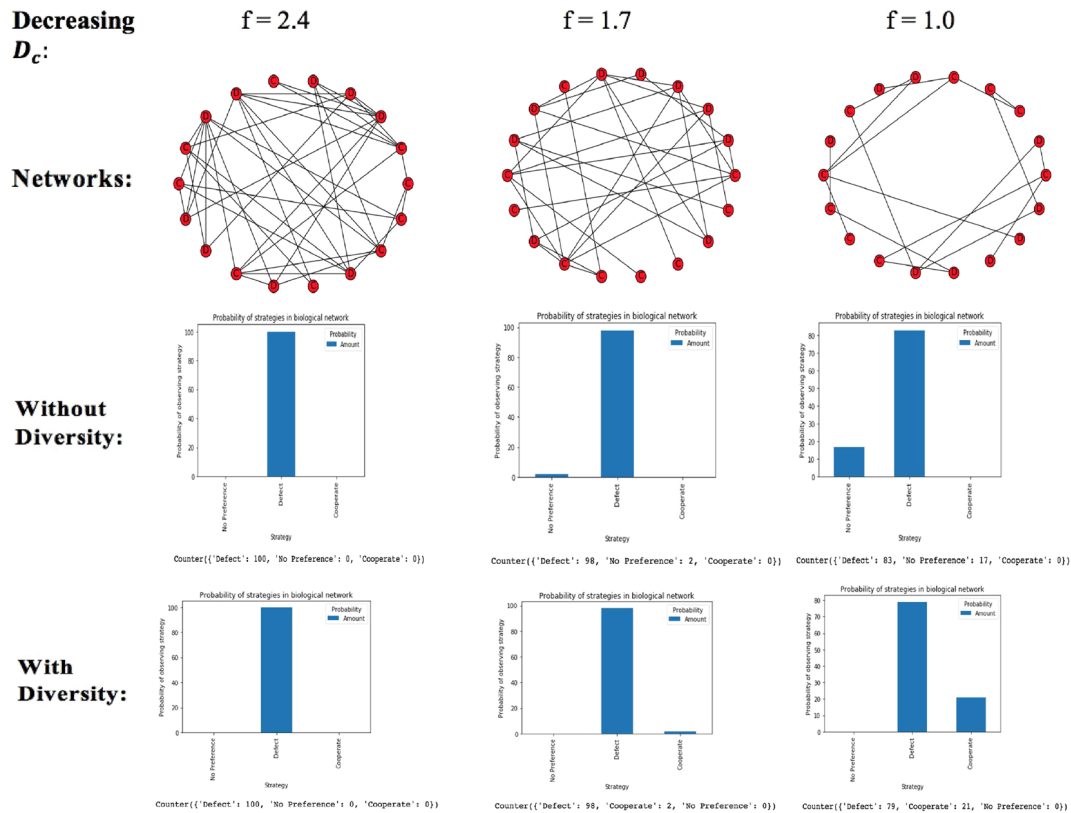

**Figure 4.** Diversity and Decreasing Density of Connectivity with no costly punishment For N = 18, f must be in the range $\frac{17}{18} \leq f \leq \frac{17}{2}$.

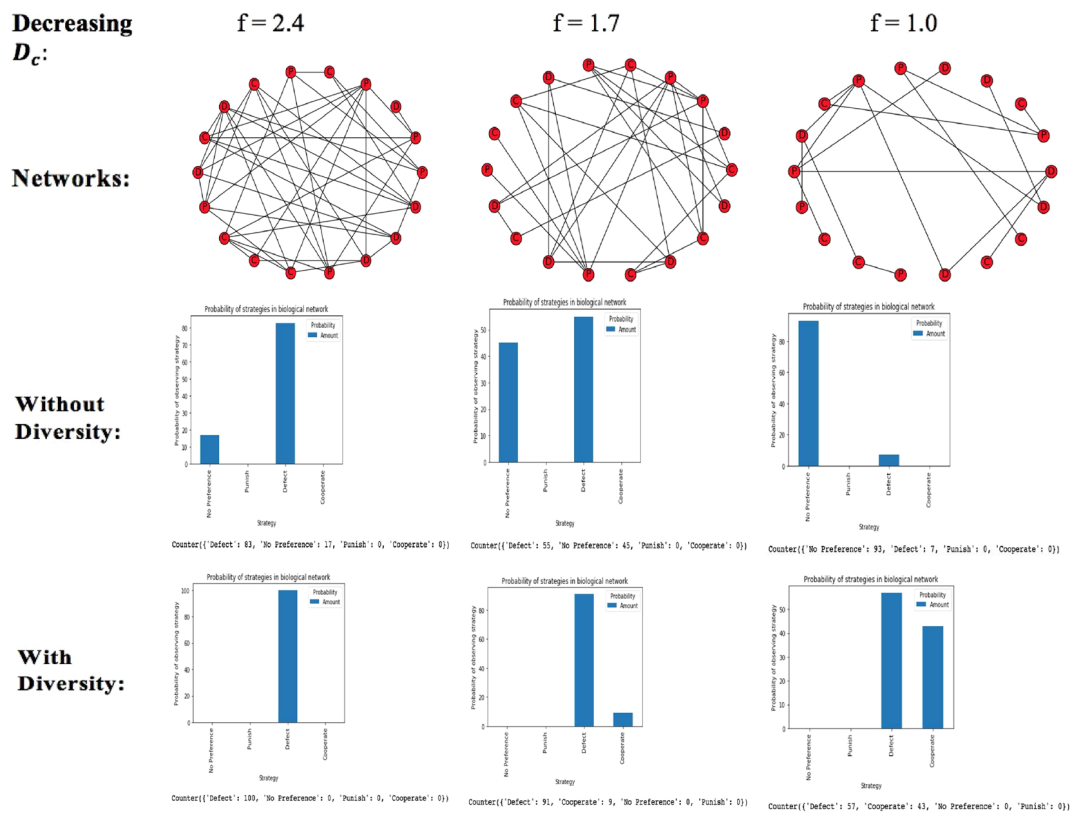

**Figure 5.** Diversity and Decreasing Density of Connectivity with costly punishment For N = 18, f must be in the range $\frac{17}{18} \leq f \leq \frac{17}{2}$.

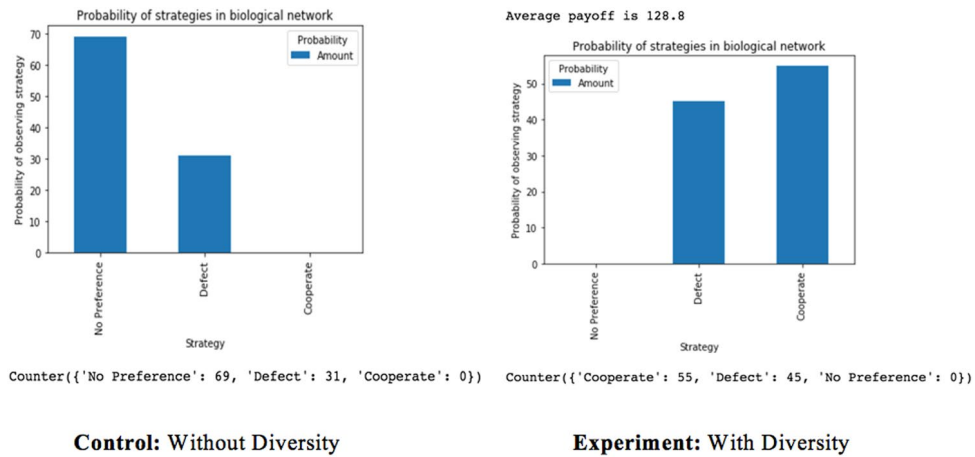

**Figure 6.** Replication of Square Lattice Experiment.

I believe the contradictory conclusions in both of our findings were not due to incorrect observations, but because of their level of analysis. Gracia-Lazaro et al[20] analyzed how the players tended towards defection in less than 71 rounds of one game even though many start out initially as cooperators. I arrived at the same finding with my agent-based model. As a matter of fact, my computational model showed that about 95.7% of the games ran had a large majority of players who defected more than cooperate in the total rounds of a game, for the experimental treatment set (with diversity/heterogeneity), but 100% of the 100 games ran had a large majority of players who tended more towards defection than cooperation in the total rounds of a game in the control treatment set (without diversity/heterogeneity). In the equilibrium/ 'No Preference' state, the few cooperators in the network (usually not more than 5 cooperators) seem to form clusters. However, on the broader scale, our model shows that introducing diversity to the network still led to the emergence of cooperation as the sole evolutionary strategy with a probability of about 55.8%. Diversity here, although promoting cooperation, also seems to enhance defection a bit. This enhancement could be due to the spatial structure of the network. The reasoning behind this speculation is due to the behavior we observe after we introduce diversity to costly punishment. We also find that both defection and cooperation are favored. Many studies have established that spatial structures favor cooperation[36–39]. In fact, this is taken to be widely accepted[30,31]. The results from this model show that costly punishment disfavors defection when employed alone, but favors both cooperation and defection when diversity is introduced. Bearing in mind that it is deceptive to translate 'disfavoring defection' as 'favoring cooperation' in the global scale, if spatial structures have an influence on defection, as we can see from this result that they do, then it appears that spatial structures do not necessarily favor cooperation in the global scale, but rather disfavor defection just like costly punishment does. It is also probable that the regular property of the random graph would affect the effect of diversity in the network. Given that Gracia-Lazaro et al[20] considered one game of less than 71 rounds, it is not surprising, from the probabilities (almost 96%), that this could give a naive understanding of the interactions in these networks and their eventual outcome when one is considering these influences at the local scale. Biological networks are complex systems, just like human societies, and we can often be misled if we do not consider the global scale of these systems.

## Conclusion
This study shows three important findings that would hopefully resolve the controversy in the literature on the impacts of costly punishment and diversity in a biological network and address the question of evolutionary theorists on how to promote large-scale cooperation. Firstly, introducing diversity to costly punishment favors both cooperation and defection, but not vice versa; whereas, introducing costly punishment to diversity disfavors defection but favors cooperation. Secondly, costly punishment, introduced alone, disfavors defection but decreases average payoff; however, decreasing the density of connectivity when there is no costly punishment applied increases the average payoff. Lastly, a synergy of diversity and decreasing density of connectivity favors cooperation in a biological network. This study needs to be extended to actual human experiments to test these observations.

## Method Summary
An agent-based model was designed using Python. The algorithm was based on the "imitate-the-best" strategy, where players imitate the strategy of the neighbor with the highest payoff given that the imitated payoff is greater than theirs; the payoff matrix was calculated following Dreber et al[23]. Three agent-based models were designed for this paper. The first was used to analyze interactions without costly punishment (link to code: https://github.com/ivanezeigbo/Biological-Networks/blob/master/teste.py). The second was designed for biological networks that implement costly punishment (link to code: https://github.com/ivanezeigbo/Biological-Networks/blob/master/punisher.py), and the last was designed to replicate the experiment of Gracia-Lazaro et al[20] using an agent-based model (link to code: https://github.com/ivanezeigbo/Biological-Networks/blob/master/graciatesing.py). The first two are random games while the last is a random k-regular graph. Updates in these models were synchronous, and randomness was applied using Python's pseudo random generator algorithm. Initially all games start with equal number of cooperators, defectors, and punishers (for networks with costly punishment). Cooperators, defectors, and punishers (for networks with costly punishment) are randomly chosen at every start of a game. The number of connections in the network is determined by the density of connectivity which is a function of f, the multiplying factor.

### Additional Information
**Competing Interests:** The authors declare that they have no competing interests.

**Publisher's note:** Springer Nature remains neutral with regard to jurisdictional claims in published maps and institutional affiliations.

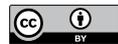 **Open Access** This article is licensed under a Creative Commons Attribution 4.0 International License, which permits use, sharing, adaptation, distribution and reproduction in any medium or format, as long as you give appropriate credit to the original author(s) and the source, provide a link to the Creative Commons license, and indicate if changes were made. The images or other third party material in this article are included in the article's Creative Commons license, unless indicated otherwise in a credit line to the material. If material is not included in the article's Creative Commons license and your intended use is not permitted by statutory regulation or exceeds the permitted use, you will need to obtain permission directly from the copyright holder. To view a copy of this license, visit http://creativecommons.org/licenses/by/4.0/.